\documentclass[aps, prl, reprint, groupedaddress,
  superscriptaddress, shortbibliography, notitlepage]{revtex4-1}

    \usepackage{graphicx} 
    \usepackage{adjustbox} 
    \usepackage{color} 
    \usepackage{enumerate} 
    \usepackage{geometry} 
    \usepackage{amsmath} 
    \usepackage{amssymb} 
    \usepackage{eurosym} 
    \usepackage[mathletters]{ucs} 
    \usepackage[utf8x]{inputenc} 
    \usepackage{fancyvrb} 
    \usepackage{grffile} 
    \usepackage{hyperref}
    \usepackage{booktabs}  
    \usepackage[lofdepth,lotdepth]{subfig}
 
    \definecolor{orange}{cmyk}{0,0.4,0.8,0.2}
    \definecolor{darkorange}{rgb}{.71,0.21,0.01}
    \definecolor{darkgreen}{rgb}{.12,.54,.11}
    \definecolor{myteal}{rgb}{.26, .44, .56}
    \definecolor{gray}{gray}{0.45}
    \definecolor{lightgray}{gray}{.95}
    \definecolor{mediumgray}{gray}{.8}
    \definecolor{inputbackground}{rgb}{.95, .95, .85}
    \definecolor{outputbackground}{rgb}{.95, .95, .95}
    \definecolor{traceback}{rgb}{1, .95, .95}
    \definecolor{red}{rgb}{.6,0,0}
    \definecolor{green}{rgb}{0,.65,0}
    \definecolor{brown}{rgb}{0.6,0.6,0}
    \definecolor{blue}{rgb}{0,.145,.698}
    \definecolor{purple}{rgb}{.698,.145,.698}
    \definecolor{cyan}{rgb}{0,.698,.698}
    \definecolor{lightgray}{gray}{0.5}
    
    \definecolor{darkgray}{gray}{0.25}
    \definecolor{lightred}{rgb}{1.0,0.39,0.28}
    \definecolor{lightgreen}{rgb}{0.48,0.99,0.0}
    \definecolor{lightblue}{rgb}{0.53,0.81,0.92}
    \definecolor{lightpurple}{rgb}{0.87,0.63,0.87}
    \definecolor{lightcyan}{rgb}{0.5,1.0,0.83}
    
    \DefineVerbatimEnvironment{Highlighting}{Verbatim}{commandchars=\\\{\}}

\makeatletter
\def\PY@reset{\let\PY@it=\relax \let\PY@bf=\relax%
    \let\PY@ul=\relax \let\PY@tc=\relax%
    \let\PY@bc=\relax \let\PY@ff=\relax}
\def\PY@tok#1{\csname PY@tok@#1\endcsname}
\def\PY@toks#1+{\ifx\relax#1\empty\else%
    \PY@tok{#1}\expandafter\PY@toks\fi}
\def\PY@do#1{\PY@bc{\PY@tc{\PY@ul{%
    \PY@it{\PY@bf{\PY@ff{#1}}}}}}}
\def\PY#1#2{\PY@reset\PY@toks#1+\relax+\PY@do{#2}}

\expandafter\def\csname PY@tok@gd\endcsname{\def\PY@tc##1{\textcolor[rgb]{0.63,0.00,0.00}{##1}}}
\expandafter\def\csname PY@tok@gu\endcsname{\let\PY@bf=\textbf\def\PY@tc##1{\textcolor[rgb]{0.50,0.00,0.50}{##1}}}
\expandafter\def\csname PY@tok@gt\endcsname{\def\PY@tc##1{\textcolor[rgb]{0.00,0.27,0.87}{##1}}}
\expandafter\def\csname PY@tok@gs\endcsname{\let\PY@bf=\textbf}
\expandafter\def\csname PY@tok@gr\endcsname{\def\PY@tc##1{\textcolor[rgb]{1.00,0.00,0.00}{##1}}}
\expandafter\def\csname PY@tok@cm\endcsname{\let\PY@it=\textit\def\PY@tc##1{\textcolor[rgb]{0.25,0.50,0.50}{##1}}}
\expandafter\def\csname PY@tok@vg\endcsname{\def\PY@tc##1{\textcolor[rgb]{0.10,0.09,0.49}{##1}}}
\expandafter\def\csname PY@tok@vi\endcsname{\def\PY@tc##1{\textcolor[rgb]{0.10,0.09,0.49}{##1}}}
\expandafter\def\csname PY@tok@mh\endcsname{\def\PY@tc##1{\textcolor[rgb]{0.40,0.40,0.40}{##1}}}
\expandafter\def\csname PY@tok@cs\endcsname{\let\PY@it=\textit\def\PY@tc##1{\textcolor[rgb]{0.25,0.50,0.50}{##1}}}
\expandafter\def\csname PY@tok@ge\endcsname{\let\PY@it=\textit}
\expandafter\def\csname PY@tok@vc\endcsname{\def\PY@tc##1{\textcolor[rgb]{0.10,0.09,0.49}{##1}}}
\expandafter\def\csname PY@tok@il\endcsname{\def\PY@tc##1{\textcolor[rgb]{0.40,0.40,0.40}{##1}}}
\expandafter\def\csname PY@tok@go\endcsname{\def\PY@tc##1{\textcolor[rgb]{0.53,0.53,0.53}{##1}}}
\expandafter\def\csname PY@tok@cp\endcsname{\def\PY@tc##1{\textcolor[rgb]{0.74,0.48,0.00}{##1}}}
\expandafter\def\csname PY@tok@gi\endcsname{\def\PY@tc##1{\textcolor[rgb]{0.00,0.63,0.00}{##1}}}
\expandafter\def\csname PY@tok@gh\endcsname{\let\PY@bf=\textbf\def\PY@tc##1{\textcolor[rgb]{0.00,0.00,0.50}{##1}}}
\expandafter\def\csname PY@tok@ni\endcsname{\let\PY@bf=\textbf\def\PY@tc##1{\textcolor[rgb]{0.60,0.60,0.60}{##1}}}
\expandafter\def\csname PY@tok@nl\endcsname{\def\PY@tc##1{\textcolor[rgb]{0.63,0.63,0.00}{##1}}}
\expandafter\def\csname PY@tok@nn\endcsname{\let\PY@bf=\textbf\def\PY@tc##1{\textcolor[rgb]{0.00,0.00,1.00}{##1}}}
\expandafter\def\csname PY@tok@no\endcsname{\def\PY@tc##1{\textcolor[rgb]{0.53,0.00,0.00}{##1}}}
\expandafter\def\csname PY@tok@na\endcsname{\def\PY@tc##1{\textcolor[rgb]{0.49,0.56,0.16}{##1}}}

\expandafter\def\csname PY@tok@nb\endcsname{\def\PY@tc##1{\textcolor[rgb]{0.00,0.50,0.00}{##1}}}
\expandafter\def\csname PY@tok@nc\endcsname{\let\PY@bf=\textbf\def\PY@tc##1{\textcolor[rgb]{0.00,0.00,1.00}{##1}}}
\expandafter\def\csname PY@tok@nd\endcsname{\def\PY@tc##1{\textcolor[rgb]{0.67,0.13,1.00}{##1}}}
\expandafter\def\csname PY@tok@ne\endcsname{\let\PY@bf=\textbf\def\PY@tc##1{\textcolor[rgb]{0.82,0.25,0.23}{##1}}}
\expandafter\def\csname PY@tok@nf\endcsname{\def\PY@tc##1{\textcolor[rgb]{0.00,0.00,1.00}{##1}}}
\expandafter\def\csname PY@tok@si\endcsname{\let\PY@bf=\textbf\def\PY@tc##1{\textcolor[rgb]{0.73,0.40,0.53}{##1}}}
\expandafter\def\csname PY@tok@s2\endcsname{\def\PY@tc##1{\textcolor[rgb]{0.73,0.13,0.13}{##1}}}
\expandafter\def\csname PY@tok@nt\endcsname{\let\PY@bf=\textbf\def\PY@tc##1{\textcolor[rgb]{0.00,0.50,0.00}{##1}}}
\expandafter\def\csname PY@tok@nv\endcsname{\def\PY@tc##1{\textcolor[rgb]{0.10,0.09,0.49}{##1}}}
\expandafter\def\csname PY@tok@s1\endcsname{\def\PY@tc##1{\textcolor[rgb]{0.73,0.13,0.13}{##1}}}
\expandafter\def\csname PY@tok@ch\endcsname{\let\PY@it=\textit\def\PY@tc##1{\textcolor[rgb]{0.25,0.50,0.50}{##1}}}
\expandafter\def\csname PY@tok@m\endcsname{\def\PY@tc##1{\textcolor[rgb]{0.40,0.40,0.40}{##1}}}
\expandafter\def\csname PY@tok@gp\endcsname{\let\PY@bf=\textbf\def\PY@tc##1{\textcolor[rgb]{0.00,0.00,0.50}{##1}}}
\expandafter\def\csname PY@tok@sh\endcsname{\def\PY@tc##1{\textcolor[rgb]{0.73,0.13,0.13}{##1}}}
\expandafter\def\csname PY@tok@ow\endcsname{\let\PY@bf=\textbf\def\PY@tc##1{\textcolor[rgb]{0.67,0.13,1.00}{##1}}}
\expandafter\def\csname PY@tok@sx\endcsname{\def\PY@tc##1{\textcolor[rgb]{0.00,0.50,0.00}{##1}}}
\expandafter\def\csname PY@tok@bp\endcsname{\def\PY@tc##1{\textcolor[rgb]{0.00,0.50,0.00}{##1}}}
\expandafter\def\csname PY@tok@c1\endcsname{\let\PY@it=\textit\def\PY@tc##1{\textcolor[rgb]{0.25,0.50,0.50}{##1}}}
\expandafter\def\csname PY@tok@o\endcsname{\def\PY@tc##1{\textcolor[rgb]{0.40,0.40,0.40}{##1}}}
\expandafter\def\csname PY@tok@kc\endcsname{\let\PY@bf=\textbf\def\PY@tc##1{\textcolor[rgb]{0.00,0.50,0.00}{##1}}}
\expandafter\def\csname PY@tok@c\endcsname{\let\PY@it=\textit\def\PY@tc##1{\textcolor[rgb]{0.25,0.50,0.50}{##1}}}
\expandafter\def\csname PY@tok@mf\endcsname{\def\PY@tc##1{\textcolor[rgb]{0.40,0.40,0.40}{##1}}}
\expandafter\def\csname PY@tok@err\endcsname{\def\PY@bc##1{\setlength{\fboxsep}{0pt}\fcolorbox[rgb]{1.00,0.00,0.00}{1,1,1}{\strut ##1}}}
\expandafter\def\csname PY@tok@mb\endcsname{\def\PY@tc##1{\textcolor[rgb]{0.40,0.40,0.40}{##1}}}
\expandafter\def\csname PY@tok@ss\endcsname{\def\PY@tc##1{\textcolor[rgb]{0.10,0.09,0.49}{##1}}}
\expandafter\def\csname PY@tok@sr\endcsname{\def\PY@tc##1{\textcolor[rgb]{0.73,0.40,0.53}{##1}}}
\expandafter\def\csname PY@tok@mo\endcsname{\def\PY@tc##1{\textcolor[rgb]{0.40,0.40,0.40}{##1}}}
\expandafter\def\csname PY@tok@kd\endcsname{\let\PY@bf=\textbf\def\PY@tc##1{\textcolor[rgb]{0.00,0.50,0.00}{##1}}}
\expandafter\def\csname PY@tok@mi\endcsname{\def\PY@tc##1{\textcolor[rgb]{0.40,0.40,0.40}{##1}}}
\expandafter\def\csname PY@tok@kn\endcsname{\let\PY@bf=\textbf\def\PY@tc##1{\textcolor[rgb]{0.00,0.50,0.00}{##1}}}
\expandafter\def\csname PY@tok@cpf\endcsname{\let\PY@it=\textit\def\PY@tc##1{\textcolor[rgb]{0.25,0.50,0.50}{##1}}}
\expandafter\def\csname PY@tok@kr\endcsname{\let\PY@bf=\textbf\def\PY@tc##1{\textcolor[rgb]{0.00,0.50,0.00}{##1}}}
\expandafter\def\csname PY@tok@s\endcsname{\def\PY@tc##1{\textcolor[rgb]{0.73,0.13,0.13}{##1}}}
\expandafter\def\csname PY@tok@kp\endcsname{\def\PY@tc##1{\textcolor[rgb]{0.00,0.50,0.00}{##1}}}
\expandafter\def\csname PY@tok@w\endcsname{\def\PY@tc##1{\textcolor[rgb]{0.73,0.73,0.73}{##1}}}
\expandafter\def\csname PY@tok@kt\endcsname{\def\PY@tc##1{\textcolor[rgb]{0.69,0.00,0.25}{##1}}}
\expandafter\def\csname PY@tok@sc\endcsname{\def\PY@tc##1{\textcolor[rgb]{0.73,0.13,0.13}{##1}}}
\expandafter\def\csname PY@tok@sb\endcsname{\def\PY@tc##1{\textcolor[rgb]{0.73,0.13,0.13}{##1}}}
\expandafter\def\csname PY@tok@k\endcsname{\let\PY@bf=\textbf\def\PY@tc##1{\textcolor[rgb]{0.00,0.50,0.00}{##1}}}
\expandafter\def\csname PY@tok@se\endcsname{\let\PY@bf=\textbf\def\PY@tc##1{\textcolor[rgb]{0.73,0.40,0.13}{##1}}}
\expandafter\def\csname PY@tok@sd\endcsname{\let\PY@it=\textit\def\PY@tc##1{\textcolor[rgb]{0.73,0.13,0.13}{##1}}}

\makeatother

    \definecolor{incolor}{rgb}{0.0, 0.0, 0.5}
    \definecolor{outcolor}{rgb}{0.545, 0.0, 0.0}

    \hypersetup{
      breaklinks=true,  
      colorlinks=true,
      urlcolor=blue,
      linkcolor=darkorange,
      citecolor=darkgreen,
      }
    
\begin{document}

\title{Molecular Simulation of Thermo-osmotic slip}
\author{Raman Ganti}
\affiliation{Department of Chemistry, University of Cambridge, Lensfield Road, Cambridge CB2 1EW, UK}
\author{Yawei Liu}
\affiliation{Beijing University of Chemical Technology, Beijing, P. R. China}
\author{Daan Frenkel}
\thanks{Corresponding author}
\email{df246@cam.ac.uk}
\affiliation{Department of Chemistry, University of Cambridge, Lensfield Road, Cambridge CB2 1EW, UK}
	
\date{\today}

\begin{abstract}

Thermo-osmotic slip -- the flow induced by a thermal gradient along a surface -- is a well-known phenomenon, but curiously there is a lack of robust molecular-simulation techniques to predict its magnitude. Here, we compare three different molecular simulation techniques to compute the thermo-osmotic slip at a simple solid-fluid interface. Although we do not expect the different approaches to be in perfect agreement, we find that the differences are barely significant for a range of different physical conditions, suggesting that practical molecular simulations of thermo-osmotic slip are feasible.

\end{abstract}
 
\maketitle   

Thermo-osmosis and thermophoresis are phenomena of great practical interest in the context of  non-isothermal hydrodynamics~\cite{morthomas2009thermophoresis,ajdari2006giant}, non-equilibrium thermodynamics~\cite{dhont2007thermodiffusion}, thermophoresis~\cite{anderson1989colloid,piazza2008thermophoresis,wurger2010thermal}, and the propulsion of active matter~\cite{golestanian2007designing}. Thermo-osmosis is usually described phenomenologically as the induced slippage  of fluid along an interface, due to an applied temperature gradient. Phoretic motion is driven by the  interfacial stresses induced by a temperature gradient in a microscopic boundary region of thickness $\lambda$, where the properties of the solvent are influenced by the interactions with the surface (or interface)~\cite{anderson1989colloid,piazza2008thermophoresis,wurger2010thermal}.

Clearly, it would be  useful to be able to predict thermophoretic slip on the basis of a molecular description of the solid-liquid interface. However, in practice this is not simple because much of the existing theoretical framework is couched in terms that assume the validity of a local continuum theory (e.g. Debye-H\"{u}ckel plus the (Navier-)Stokes equation). Yet, crucially, near an interface, a continuum description  of the structure or dynamics of a liquid is not allowed. More ominously, the definition of the stress in a liquid is not unique. This non-uniqueness has no effect on the computed value of, say, the liquid-liquid surface tension~\cite{schofield1982statistical}, but it could affect the prediction of phoretic flows, where the {\em local} value of the stress gradient is what drives the flow. In this paper, we consider this problem and explore various `microscopic' methods to predict thermo-osmotic slippage in a simple model system.

The `classical' approach to predict thermophoretic slippage is based on Onsager's reciprocity relations (see Ref.~\cite{degroot}). Derjaguin~\cite{derjaguin_surface}  used Onsager's Linear Non-Equilibrium Thermodynamics (LNET) approach to derive an expression for thermo-osmotic slip. Derjaguin's approach exploits the relation between the flow caused by a temperature gradient and the excess heat flux due to hydrodynamic flow, resulting in the following equation for thermo-osmotic slip:
\begin{equation}
\label{eq:derjaguin}
v_{s} = -\frac{2}{\eta}\int_{0}^{\infty} \mathrm{d}z \,z\Delta h(z)\frac{\nabla T}{T},
\end{equation}
where $\Delta h(z)$ is the excess enthalpy density at a height $z$ above the surface and $\eta$ is the viscosity. The difficulty with this expression is that there is some ambiguity in the microscopic definition of the local excess enthalpy $\Delta h(z)$, a quantity that is also not easy to probe in experiments~\cite{anderson1989colloid}. Moreover, Eq~\eqref{eq:derjaguin} assumes that the (Navier-)Stokes equation with constant viscosity holds very close to a surface. A correct, microscopic description would not make such a continuum assumption. For these reasons, molecular simulations should be the method of choice to predict and study thermo-osmotic flows.  

To place the various descriptions of thermo-osmotic slip in a broader context, we first consider the classical thermodynamic approach  to the problem, based on the assumption of Local Thermal Equilibrium (LTE).
We note that neither temperature gradients nor, for that matter,  chemical potential gradients can exert a net force on a fluid element in a bulk liquid. Mechanical forces in liquids can only be caused by body forces such as gravity, or by pressure gradients. If temperature gradients cause flow near a surface, it is only because a local pressure gradient is induced. To clarify this, we first consider the thermodynamics of the problem. Consider a temperature gradient  $+x$ direction parallel to a hard wall; the $z$ coordinate measures distance perpendicular to the wall. Starting from the Gibbs-Duhem relation for an $n$-component mixture, we write
\begin{equation}\label{eq:VdP}
VdP = \sum_{i = 1}^{n} N_{i}d\mu_{i} + SdT.
\end{equation}
The Gibbs-Duhem equation makes use of the fact that the system is homogeneous. A stratified system in equilibrium, is homogeneous in the directions parallel to the stratification, but not perpendicular to it. Hence, here and in what follows, the `pressure' $P$ refers to a component of the pressure tensor parallel to the surface (e.g. $P_{xx}$).  Dividing Eq~\eqref{eq:VdP} by $V$ and differentiating with respect to $x$ gives the following expression:
\begin{equation}
\label{eq:GB_pressure_gradient}
\frac{\partial P}{\partial x} = \left(\sum_{i = 1}^{n} \rho_{i}\frac{\partial \mu_{i}}{\partial T} + \frac{S}{V}\right)\frac{\partial T}{\partial x},
\end{equation}
where $\rho_i$ is the number density of species $i$. In the bulk, the pressure is equalised quickly and the fluid reaches hydrostatic equilibrium. Since the bulk pressure is constant, Eq~\eqref{eq:GB_pressure_gradient} reduces to
\begin{equation}
\label{eq:specific_entropy}
\left(\frac{\partial \mu_{i}}{\partial T}\right)_{P} = -s^{B}_{i}
\end{equation}
using $S = \sum_i N_i s_i$, where $s_{i}$ is the specific entropy of species $i$ and the superscript $B$ denotes a bulk quantity. We now write a similar expression for the pressure gradient at a position $z$ above the surface
\begin{equation}
\label{eq:gradP_at_z}
\frac{\partial P(z)}{\partial x} = \left(\sum_{i = 1}^{n} \rho_{i}(z)\frac{\partial \mu_{i}}{\partial T} + \frac{S(z)}{V}\right)\frac{\partial T}{\partial x}.
\end{equation}
We assume that there are no gradients of $\mu_{i}$ and $T$ perpendicular to the surface. Using Eq~\eqref{eq:specific_entropy}, Eq~\eqref{eq:gradP_at_z} can be rewritten as
\begin{equation}
\label{eq:sub_bulk}
\frac{\partial P(z)}{\partial x} = \left(-\sum_{i = 1}^{n} \rho_{i}(z)s^{B}_{i} + \sum_{i = 1}^{n} \rho_{i}(z)s_{i}(z)\right)\frac{\partial T}{\partial x}.
\end{equation}
Eq~\eqref{eq:sub_bulk} can be simplified by noting that the expression in brackets is the difference between the specific
entropy at position $z$ and the bulk specific entropy. Since $\mu_{i}$ and $T$ do not depend on $z$, $\mu_{i} = h_{i} - Ts_{i}$ can be used to rewrite Eq~\eqref{eq:sub_bulk} as
\begin{eqnarray}
\label{eq:gradP_enthalpy}
\frac{\partial P(z)}{\partial x} &= \left(\frac{\sum_{i = 1}^{n} \rho_{i}(z)[h_{i}(z) - h^{B}_{i}]}{T}\right)\frac{\partial T}{\partial x} \\
&= \left(\frac{\Delta{h(z)}}{T}\right)\frac{\partial T}{\partial x},
\end{eqnarray}
where \(\Delta{h(z)}\) is the excess enthalpy density at a distance \(z\) from the surface. To compute the flow velocity, we combine the expression for the pressure gradient with the linearised (Navier-)Stokes equation
\begin{equation}
\label{Navier-Stokes}
\eta\left(\frac{\partial^{2}v_{x}(z)}{\partial z^{2}}\right) = \left(\frac{\partial P(x,z)}{\partial x}\right).
\end{equation}
Assuming no surface slip, this equation can be integrated twice to yield 
\begin{equation}
\label{eq:slip}
v_{s} = -\frac{1}{\eta}\int_{0}^{\infty}\mathrm{d}z\;z\left(\frac{\Delta{h(z)}}{T}\right)\frac{\partial T}{\partial x}.
\end{equation}
This is equivalent to Derjaguin's expression in Eq~\eqref{eq:derjaguin} in the case of a single surface. The advantage of the LTE approach is that Eq~\eqref{eq:gradP_enthalpy} demonstrates the relationship between the external temperature gradient and the local pressure gradient that induces thermo-osmotic slip. The magnitude of this coupling is determined exclusively by the excess surface enthalpy.
\begin{figure}
\begin{center}
\adjustimage{max size={0.5\linewidth}{0.5\paperheight}}{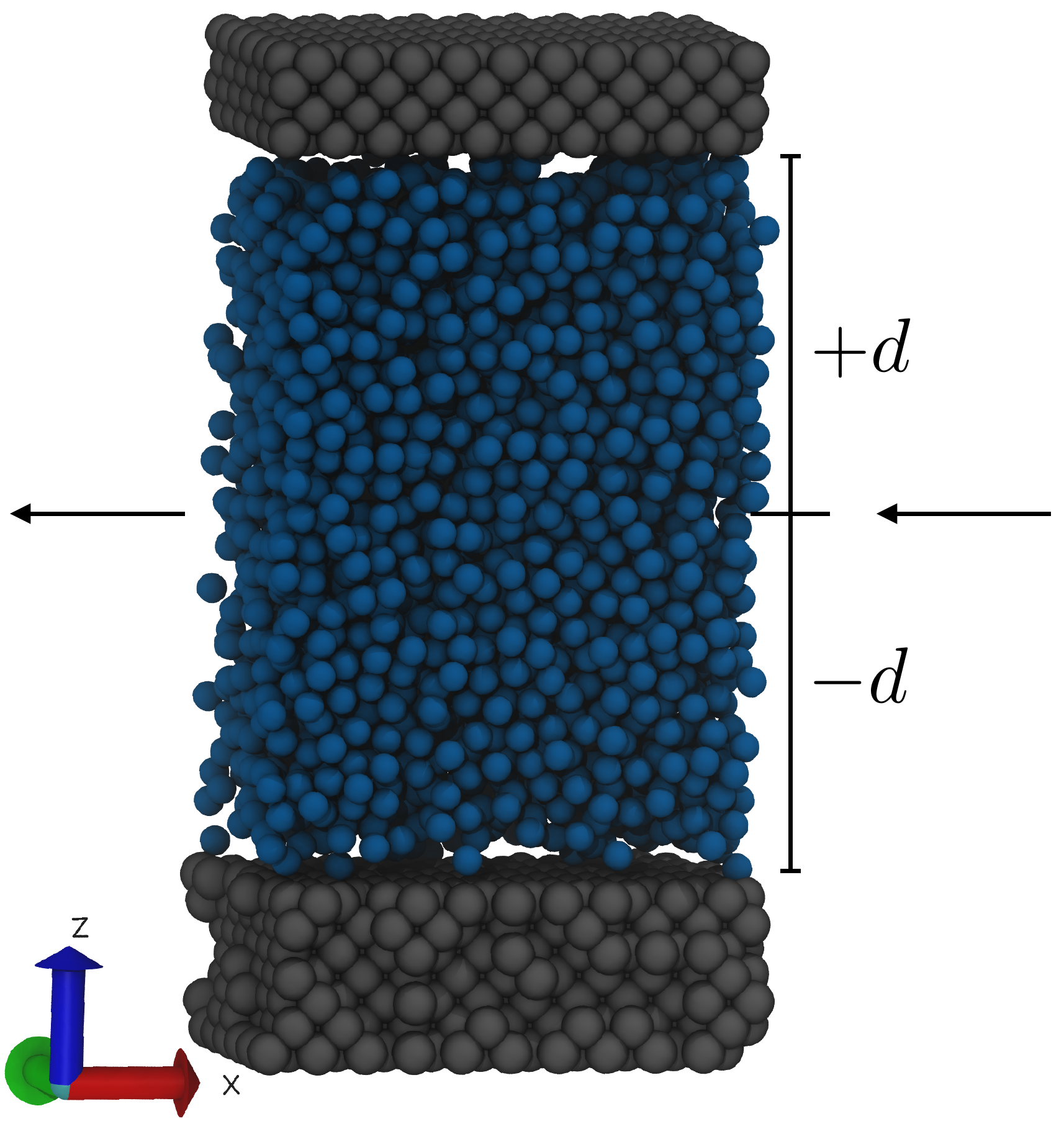}
\caption{Atomic fluid (blue) interacting with solid walls (grey) in a slit pore. $2d$ is the gap width.\label{fig:md_slit}}
\end{center}
\end{figure}
To relate our expression to Derjaguin's LNET approach, consider the slit pore as depicted in Fig. \ref{fig:md_slit}. Following Derjaguin, a pressure and temperature gradient is maintained across the slit. Fluid flows in the $-x$ direction as depicted by the arrows. For a  one-component fluid, the rate of entropy production can be written as
\begin{equation}
\label{eq:entropy_production}
T\rho\dot{s} = -v_{x}\nabla{P} - Q_{x}\frac{\nabla T}{T},
\end{equation}
where \(v_{x}\) is the fluid velocity (\(m/s\)) and \(Q_{x}\) is the heat flux (\(J/(m^{2}\cdot s)\)). Eq~\eqref{eq:entropy_production} implies the following phenomenological equations:
\begin{eqnarray}
\label{eq:phenomenological}
v_{x} = -\beta_{11}\nabla{P} -\beta_{12}\frac{\nabla T}{T} \\
Q_{x} = -\beta_{21}\nabla{P} -\beta_{22}\frac{\nabla T}{T}.
\end{eqnarray}
Consider the isothermal heat flux across the pore in Fig. \ref{fig:md_slit}
\begin{equation}
\label{eq:heat_flux}
Q_{x} = h^{B}v_{x} + \frac{1}{2d}\int_{-d}^{+d}\Delta{h(z)}v_{x}(z)\mathrm{d}z.
\end{equation}
The first term on the right-hand side of Eq~\eqref{eq:heat_flux} is the advective term and equivalent to the bulk heat content of the fluid that is transported across the pore. The second term is the transport of excess heat in the boundary layers. The second term is what determines the strength of thermo-osmotic slip~\cite{degroot}. Therefore, we can write
\begin{equation}
\label{eq:beta21}
\beta_{21} = -\left(\frac{Q_{x} - h^{B}v_{x}}{\nabla{P}}\right)_{T}.
\end{equation}
$\beta_{21}$ defined here is conventionally known as the `mechano-caloric' coefficient. Using Derjaguin's approximation that the velocity profile is linear in the boundary layer $v_{x}(z) = -2dz\nabla{P}/\eta$ and considering only the bottom wall, $\beta_{21}$ can also be expressed as
\begin{equation}
\beta_{21} = \frac{1}{\eta}\int_{0}^{\infty}\mathrm{d}z\;z\Delta{h(z)}.
\end{equation}
By considering the isobaric mass flux in Eq~\eqref{eq:phenomenological}, we can write
\begin{equation}
\label{eq:beta12}
\beta_{12} = -\left(\frac{v_{x}}{(\nabla{T}/T)}\right)_{P}.
\end{equation}
Substituting Eq~\eqref{eq:slip} for $v_{x}$ in Eq~\eqref{eq:beta12} immediately shows $\beta_{12} = \beta_{21}$ as expected. This provides the link between our LTE and Derjaguin's LNET approach.

The usual definition of the `slip' velocity is the extrapolated velocity at the interface, where the fluid density approaches zero. For a thin boundary layer, the  slip velocity is equal to the fluid velocity in the bulk just outside the boundary layer. $\beta_{12}$ in Eq~\eqref{eq:beta12} is the `thermo-osmosis coefficient.'

An alternative approach to the LTE route is to compute the thermo-osmotic slip, using a mechanical route, i.e. by computing the force on a volume element directly from the gradient of the microscopic stress. Such an approach has been used by Han~\cite{han2005}, but it could be problematic due to the non-uniqueness of the definition of the microscopic stress.
We start with the relation between the stress gradient and $f_{x}(z)$ the force per unit volume on a fluid element at a distance $z$ from the surface. Rather than computing the stress gradient in a non-equilibrium simulation, we use the fact that $P_{xx}$ depends on $x$, only through $T$. Hence,
\begin{equation}
\label{eq:body_force}
f_{x}(z) = - \left(\frac{P^{eq,T_2}_{xx}(x,z) - P^{eq,T_1}_{xx}(x,z)}{T_{2} - T_{1}}\right)\left(\frac{\partial T}{\partial x}\right),
\end{equation}
where the superscript $eq$ denotes equilibrium calculations that are both carried out at the same bulk pressure. With this method, $\Delta{P}/\Delta{T}$ is determined, and for any $\partial T/\partial x$, $f_{x}(z)$ can be computed via Eq~\eqref{eq:body_force}. The thermo-osmotic force per particle $f^{P}_{x}(z) = f_{x}(z)/\rho_{ave}(z)$ where $\rho_{ave}(z) = (\rho_{T_1}(z) + \rho_{T_2}(z))/2$. To compute the thermo-osmotic flow, we now carry out a second simulation at the average temperature $T_{ave} = (T_{1} + T_{2})/2$, where we apply the local body force $f_x(z)$. The resulting slip velocity and therefore $\beta_{12}$ can then be computed. Note that in this approach, we make no continuum assumptions.

The calculation as described above is complicated by the fact that  the pressure tensor in an inhomogeneous fluid is not unique~\cite{hafskjold2002microscopic,rowlinson2013molecular}. Irving and Kirkwood (IK)~\cite{irving1950} proposed an expression by integrating the total momentum flux acting across a virtual surface element. This approach gives the appropriate mechanical force balance normal to the interface. However, as argued by Schofield and Henderson~\cite{schofield1982statistical}, the definition of the pressure tensor is not unique since any term with a vanishing divergence can be added without changing the momentum flux. All common definitions do, however, yield the correct surface tension.

In a simulation,  we need to know the thermo-osmotic force acting on atoms,  as opposed to the force on the fictitious surface of a volume element.  This would suggest that the atom-based virial (V) expression for pressure might be preferable. 

In order to determine if the choice of the pressure affects the computed thermo-osmosis coefficient, we computed $P_{xx}$ in Eq~\eqref{eq:body_force} using both the V and IK expressions. The V pressure is given by~\cite{hansen1990theory}
\begin{equation}
\label{eq:virial_pressure}
P^{Vir}_{xx}(z) = \langle\rho(z)\rangle k_{B}T - \frac{1}{V(z)}\left\langle\frac{1}{2}\sum^{N(z)}_{i}\sum_{j\ne i}\frac{x_{ij}^{2}}{r_{ij}}\phi'(r_{ij})\right\rangle.
\end{equation}
The IK pressure is computed using \cite{walton1983pressure}
\begin{equation}
\begin{split}
\label{eq:ik_pressure}
P^{IK}_{xx}(z) &= \langle\rho(z)\rangle k_{B}T \\ &- \frac{1}{2A}\left\langle\sum^{N}_{i}\sum_{j\ne i}\frac{x_{ij}^{2}}{r_{ij}}\frac{\phi'(r_{ij})}{|z_{ij}|}\Theta\left(\frac{z - z_{i}}{z_{ij}}\right)\Theta\left(\frac{z_{j} - z}{z_{ij}}\right)\right\rangle.
\end{split}
\end{equation}
In addition to these `mechanical' expressions for the thermo-osmotic slip, consider the right-hand side of Eq~\eqref{eq:gradP_enthalpy}. We express the local specific enthalpy as
\begin{equation}
\label{eq:local_enthalpy}
h(z) = u(z) + \frac{P^{Vir}_{xx}(z)}{\rho(z)},
\end{equation}
where $u$ is the specific internal energy. In Eq~\eqref{eq:local_enthalpy}, we have made explicit that the pressure that enters into the expression for the local enthalpy must be the component that is parallel to the surface, as argued below Eq~\eqref{eq:VdP}.
The body force on a fluid element at a height $z$ above the surface is then given by
\begin{equation}
\label{eq:exc_enthalpy_force}
f_{x}(z) = -\frac{\rho(z)(h(z) - h^{B})}{T}\left(\frac{\partial T}{\partial x}\right).
\end{equation}
Eq~\eqref{eq:exc_enthalpy_force} can be computed in a simulation thermostatted at $T_{ave}$ and applied as a body force in the same vein as Eq~\eqref{eq:body_force}.

We compare the above calculations of the slip coefficient with the result for $\beta_{12}$ that follows from Derjaguin's approach based on the Onsager reciprocity relations. In this case, a uniform pressure gradient represented by a body force is applied during a simulation at $T_{ave}$.  $\beta_{21}$ is computed via Eq~\eqref{eq:beta21} (see Supplemental Material S2). The LTE approaches for computing $\beta_{12}$ and the `Derjaguin' method for computing $\beta_{21}$ should be equivalent if the temperature and pressure gradients are small enough to ensure that the resulting response is linear.

All Molecular Dynamics simulations reported here were performed using the LAMMPS package \cite{plimpton1995fast}. 
The simulation setup is depicted in Fig. \ref{fig:md_slit}.
 The system consists of a Lennard-Jones fluid of $N = 2640$ atoms interacting with a solid atomic wall.
\begin{equation}
V_{\mathrm{trunc}}(r)=\begin{cases}
    4\epsilon \left[\left(\frac{\sigma}{r}\right)^{12} - \left(\frac{\sigma}{r}\right)^{6}\right] - V(r_{c}) &r \le r_{c}\\
    0 &r > r_{c}.
  \end{cases}
\end{equation}
Interaction parameters are set so that \(\sigma_{\text{fluid-fluid}} = \sigma_{\text{wall-fluid}} = \sigma\) and \(\epsilon_{\text{fluid-fluid}} = \epsilon\) with \(r_c = 4\sigma\). Two different wall-fluid interactions were considered: a less attractive Lennard-Jones where $\epsilon_{\text{wall-fluid}} = 0.55\epsilon$ and a purely repulsive Weeks-Chandler-Andersen potential~\cite{weeks1971role} such that $r_c = 2^{1/6}\sigma$. Wall atoms are fixed via harmonic springs. All computed quantities are expressed in Lennard-Jones  units.

NVT dynamics with a time step $\Delta{t} = 0.001$ were run to  equilibrate the system. This was accomplished using a Nos{\'e}-Hoover thermostat for $100,000$ MD steps. For an additional $100,000$ steps, the system was barostatted at \($P$ \approx 0.122\) by applying a downward force to the top wall atoms.

\begin{figure}[h]
\includegraphics[width=1.0\columnwidth]{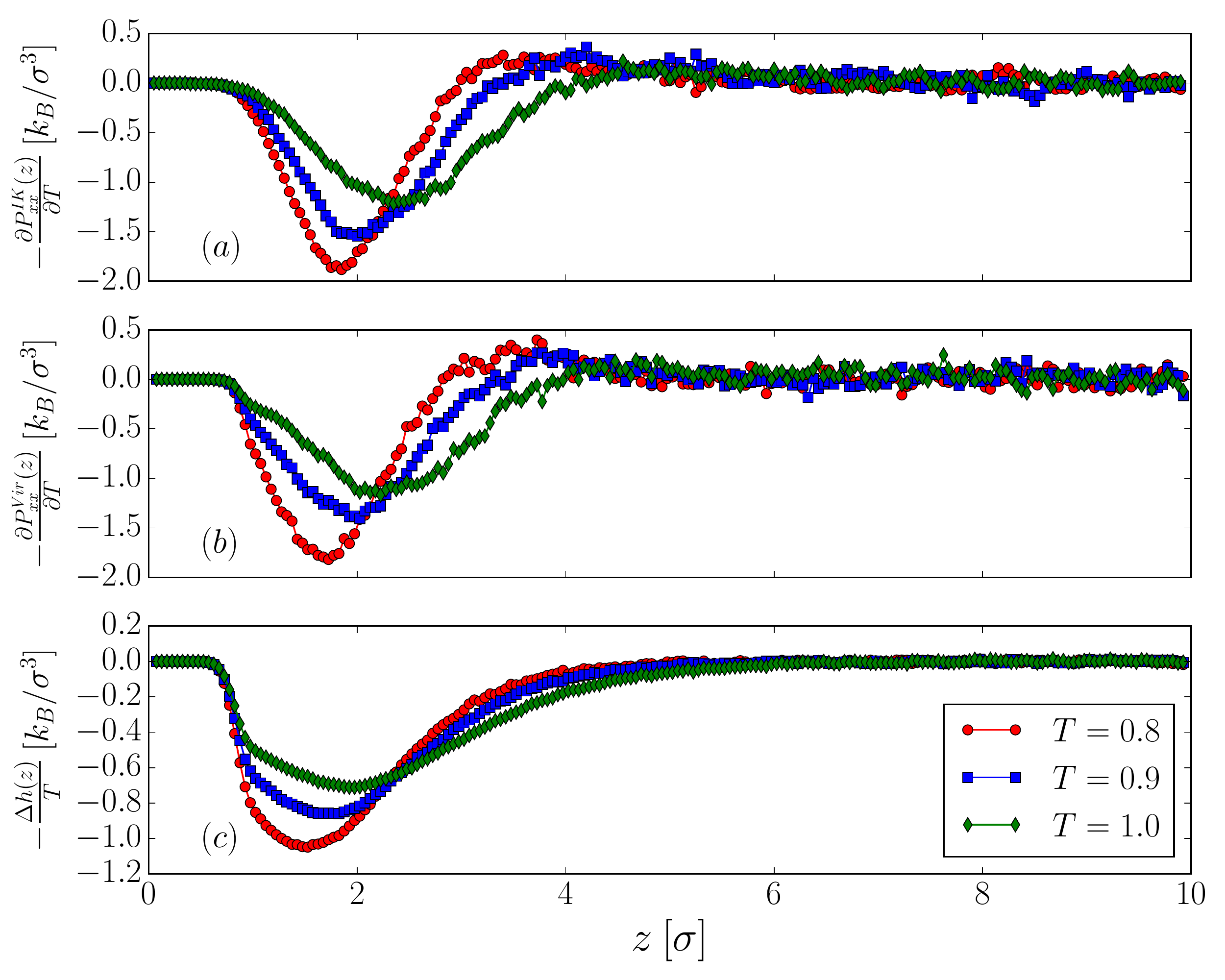}
\caption{(a) and (b) show $-\Delta{P_{xx}}/{\Delta{T}}$ e.g. the profile for $T=0.8$ is computed by taking the difference in $P_{xx}$ at $T=0.85$ and $0.75$ and dividing by $\Delta{T} = 0.1$. WCA wall-fluid interactions significantly exclude volume and thereby create a large enthalpy difference at the surface as shown in (c). The solid wall is located at $z\sim 0$. \label{fig:gradient_profiles}}
\end{figure}

The solid surface introduces local anisotropy in the pressure that vanishes in the bulk. The response of this anisotropy to a change in temperature drives thermo-osmotic slip. Using the pressure profiles (see Supplemental Material: Figs. S1(a,b)), $\Delta{P_{xx}(z)}/\Delta{T}$ was computed for the three temperatures shown in Figs. \ref{fig:gradient_profiles}(a) and \ref{fig:gradient_profiles}(b). Encouragingly, the choice of the pressure tensor makes no significant difference to the measured response.

At constant temperature, $T_{ave}$, the specific kinetic energy is uniform everywhere and therefore computed by dividing the total average kinetic energy by the number of atoms. For the same temperatures, the specific potential energy profiles were spatially averaged in $z$. Using the profiles of $P_{xx}$ (Fig. S1(a)) and density profiles (Fig. S1(c)), $\Delta{h(z)}/T$ was computed via Eqs~\eqref{eq:local_enthalpy} and \eqref{eq:exc_enthalpy_force} and shown in Fig. \ref{fig:gradient_profiles}(c).

We note that the V and IK expressions (Figs.~\ref{fig:gradient_profiles}(a), \ref{fig:gradient_profiles}(b)) and the LTE quantity Fig.~\ref{fig:gradient_profiles}(c) show similar qualitative behavior. The body force vanishes in the bulk as it should. In both cases, the profiles flatten and shift outward as the temperature is increased.

\begin{figure}[h]
\includegraphics[width=1.0\columnwidth]{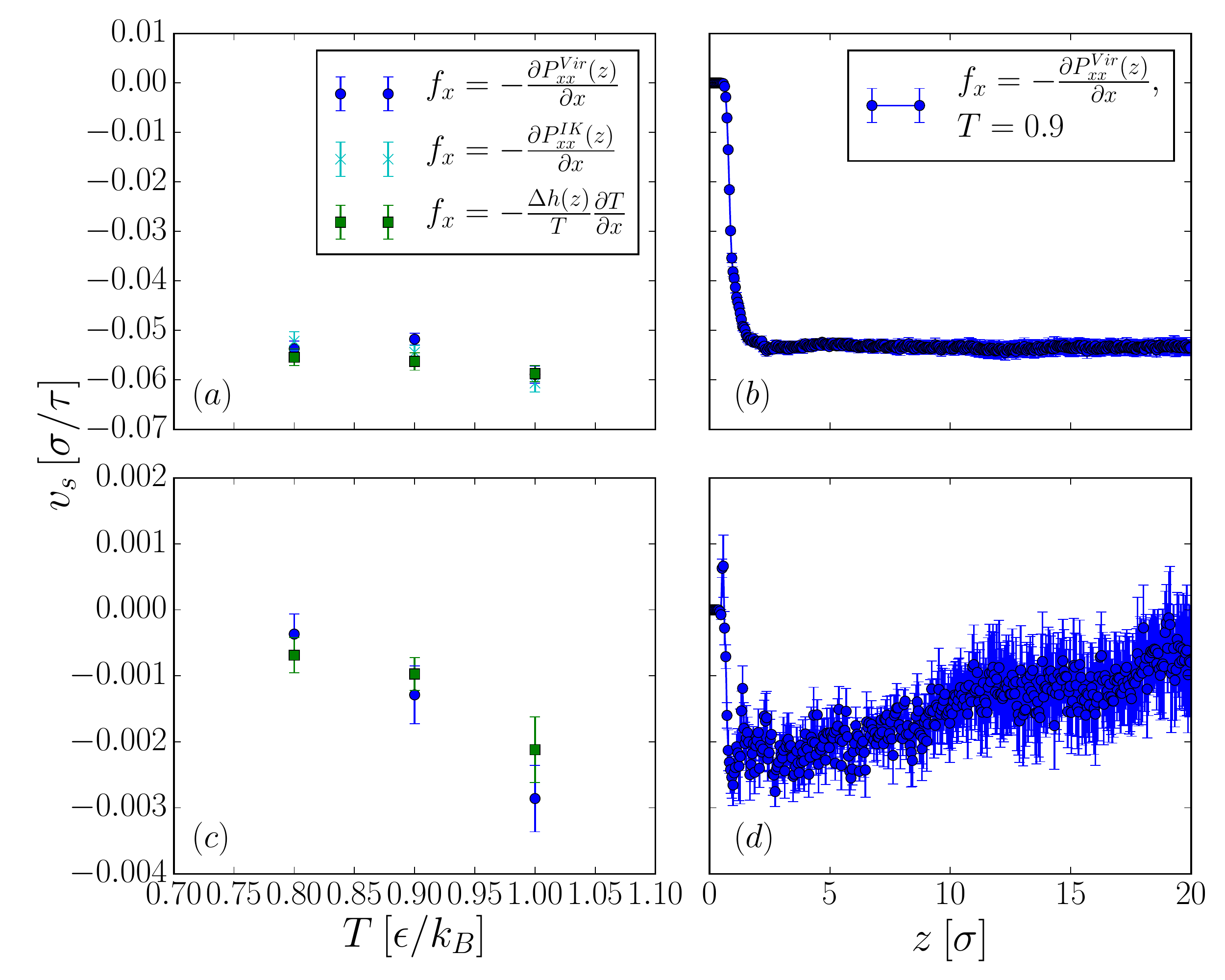}
\caption{Calculations of the slip velocity and flow profile for different wall-fluid interactions: (a,b) WCA at $\nabla{T}=0.0005$ and (c,d) Lennard-Jones ($\epsilon_{wf} = 0.55\epsilon$) at $\nabla{T}=0.003$. The slip appears to increase with temperature. This is consistent with Figs. \ref{fig:gradient_profiles}(a-c) as the pressure gradient and force due to the excess enthalpy density shift outward and thereby act on a greater number of fluid atoms. \label{fig:slip_profiles}}
\end{figure}

The body force per particle $f^{P}_{x}(z)$ can be computed by dividing the profiles in Figs. \ref{fig:gradient_profiles}(a-c) by $\rho(z)$ (Fig. S1(c)) and multiplying by a sufficiently small gradient e.g. $\partial T/\partial x = 0.0005$ for WCA wall-fluid interactions. To compute slip, non-equilibrium simulations were carried out by applying these forces to the equilibrated systems at the appropriate temperatures. To obtain reasonable statistics, forces were applied for $10^{8}$ steps until the fluid approached a steady velocity. The slip was then computed for an additional $2\times10^{8}$ steps. Figs. \ref{fig:slip_profiles}(a,c) show calculations of the slip velocity. There appears to be reasonably good agreement between the three approaches and the choice of the pressure tensor appears to make no difference. The flow profile computed at $T = 0.9$ (Fig. \ref{fig:slip_profiles}(b)) shows that for WCA wall interactions the velocity decreases monotonically indicating that the viscosity close to the surface is constant. For less attractive Lennard-Jones (Fig. \ref{fig:slip_profiles}(d)), the viscosity is clearly not constant showing significant departure from (Navier-)Stokes and Derjaguin's result (Eq~\eqref{eq:derjaguin}).

\begin{figure}[h]
\includegraphics[width=1.0\columnwidth]{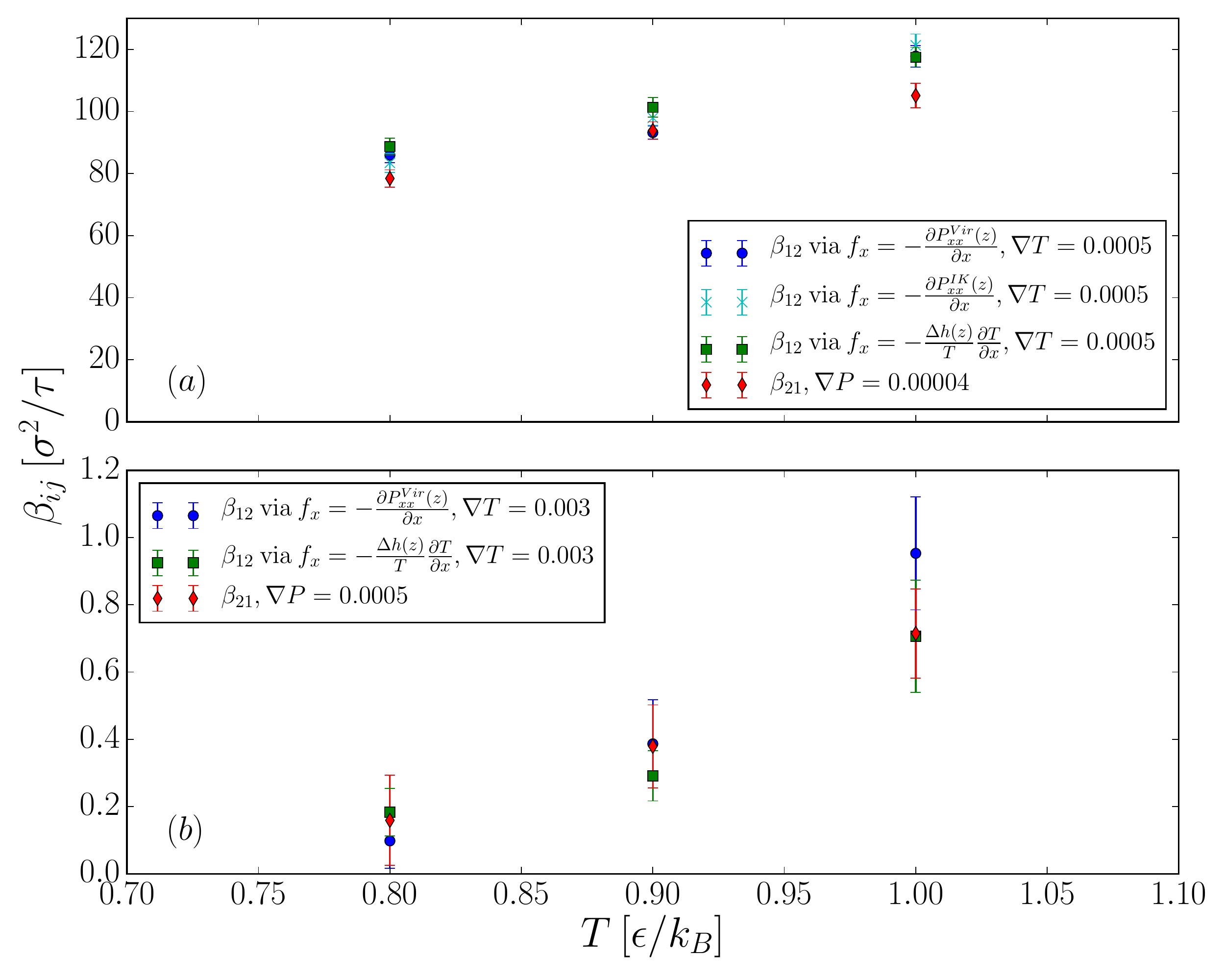}
\caption{Comparison of Onsager reciprocal relations, $\beta_{12}$ computed via our `stress gradient' and LTE approaches and $\beta_{21}$ calculated by following Derjaguin's LNET method. Thermo-osmosis coefficients are computed for (a) WCA and (b) Lennard-Jones interactions. \label{fig:beta_vs_T}}
\end{figure}

To compare our LTE approaches and Derjaguin's method (see Supplemental Material S2), $\beta_{12}$ was computed via Eq~\eqref{eq:beta12} using the slip calculations shown in Figs. \ref{fig:slip_profiles}(a,c). Fig. \ref{fig:beta_vs_T} shows a comparison of all three methods. For the range of temperatures $T\sim 0.8-1.0$, there appears to be reasonable agreement. There is some discrepancy between $\beta_{12}$ computed via Eq~\eqref{eq:exc_enthalpy_force} and $\beta_{21}$ in Fig. \ref{fig:beta_vs_T}(a). This may be due to noise in the force profile or fluctuations in the barostat. As expected, $\beta_{12}$ and $\beta_{21}$ for solely repulsive wall-fluid interactions are considerably larger than those for interactions with an attractive component. This is entirely due to the amount of slip at the surface in response to either a temperature or pressure gradient. Furthermore, both cases demonstrate an approximately linear dependence of the thermo-osmosis coefficient with respect to temperature.

In addition to the methods described above, we also attempted to compute $\beta_{21}$ using linear-response theory~\cite{luttinger1964theory}. However, no reliable results were obtained as the statistical noise overwhelmed the signal.

In summary, we have considered four different methods to compute thermo-osmotic slip on the basis of molecular simulations. The first approach is based on a computation of the thermally-induced stress gradient method, computed using equilibrium simulations and then represented as a body force in non-equilibrium simulations. We find no evidence that different choices for the pressure tensor lead to different results. In the second approach, we compute the excess enthalpy density near the wall and use a local-thermodynamics formalism to derive the body force acting on the fluid. These methods do not assume that macroscopic thermodynamics or hydrodynamics holds close to an interface. The final approach is based on Onsager's reciprocal relations, which allow us to derive thermo-osmotic slip from the excess heat flux due to a pressure gradient.

Our results are encouraging and surprising: we find that all methods yield results for the thermo-osmotic slip that do not differ significantly. Hence, the choice of the method to compute thermo-osmotic slip seems to be a matter of taste or convenience. 

We gratefully acknowledge numerous discussions with Lyd\'{e}ric Bocquet, Mike Cates, Patrick Warren, Ignacio Pagonabarraga and Benjamin Rotenberg. Additionally, we are grateful to Peter Wirnsberger for his help with the error analysis. RG gratefully acknowledges a PhD Grant from the Sackler Fund.

\clearpage
\bibliographystyle{apsrev4-1}
\bibliography{thermo_bib}
     
\end{document}